\documentclass[structabstract]{aa}  

\usepackage[dvips]{graphicx}
\usepackage{txfonts}
\usepackage{natbib}

\usepackage{ulem}
\usepackage{subfig}

\begin{document}
%
\title{The eccentricity distribution of compact binaries}

   \author{I. Kowalska
          \inst{1}
          \and
          T. Bulik\inst{1,2}
          \and
          K. Belczynski\inst{1,3}
          \and 
          M. Dominik \inst{1}
          \and 
          D. Gondek-Rosinska \inst{4,2}
          }

   \institute{Astronomical Observatory, University of Warsaw, Al Ujazdowskie 4, 00-478 Warsaw, Poland
\and Nicolaus Copernicus Astronomical Center, Bartycka 18, 00716, Warsaw, Poland
\and Dept. of Physics and Astronomy, University of Texas,  Brownsville, TX 78520, USA
\and Institute of Astronomy, University of Zielona G\'ora, ul. Lubuska 2, 65-265 Zielona G\'ora, Poland}      

   \date{Received  ; accepted  }
 
\abstract
   {The current gravitational wave detectors have reached their operational sensitivity 
and are nearing detection of compact object binaries. In the coming years, we expect 
that the Advanced LIGO/VIRGO will start taking data. At the same time, there are plans
for third generation ground-based detectors such as the Einstein Telescope, and 
 space detectors such as DECIGO. }
   {
We discuss the eccentricity distribution of inspiral compact object binaries
during they inspiral phase. We analyze the expected distributions 
of eccentricities at three frequencies that are characteristic of three future detectors: Advanced
LIGO/VIRGO (30 Hz), Einstein Telescope (3 Hz), and DECIGO (0.3 Hz).
}
   {We use the  {{\tt StarTrack}} binary population code to investigate 
the properties of the population of compact binaries in formation. We evolve their orbits until 
the point that they enter a given detector sensitivity window and analyze the eccentricity distribution at that time.}
   {We find that the eccentricities of BH-BH and BH-NS binaries are 
quite small when entering the Advanced LIGO/VIRGO detector window for all considered models 
of binary evolution.
Even in the case of the DECIGO detector, the typical eccentricities 
of BH-BH binaries are below $10^{-4}$, and the BH-NS eccentricities are smaller than $10^{-3}$.
Some fraction of NS-NS binaries may have significant eccentricities.
Within the range of considered models, we found that a fraction of between 0.2\% and 2\% 
NS-NS binaries will have an eccentricity above 0.01 for the Advanced LIGO/VIRGO detectors.
For the ET detector, this fraction is between 0.4\% and 4\%, and for the DECIGO
detector it lies between 2\% and 27\%.
}
   {}


   \maketitle
%

\section{Introduction}

As the interferometric gravitational wave detectors LIGO and VIRGO
\citep{2010CQGra..27h4006H,2006CQGra..23S..63A} reach their design sensitivities, the first detection of gravitational waves
has become more imminent. Both detectors will undergo serious improvements to increase their
sensitivity \citep{2009CQGra..26k4013S,2005CQGra..22S.461S}. 

It is therefore important to investigate the 
properties of the primary candidate sources for detection, namely compact object binaries.
There have been a number of papers dealing with 
several properties of the population of compact binaries \citep[e.g.,][]{2001A&A...376..950N,2003MNRAS.342.1169V,2004NewA....9....1D,2002ApJ...572..962S,2005ApJ...628..343P,2002MNRAS.331.1027D,2005MNRAS.363L..71D,2007ARep...51..308B,2010MNRAS.406..656K}.
In particular, \citet{2010CQGra..27q3001A} presented their estimated detection rates.
In addition, the mass spectrum \citep{2007AdSpR..39..285G},
and even spin properties \citep{2004PhRvD..70l4020S,2010CQGra..27k4007M} have been studied.
In this paper, we present yet another aspect of the 
merging compact object binary population: the distribution of the eccentricity.

From radio observations, we currently know of only six compact object binaries with merger
timescales shorter than a Hubble time, all of them NS-NS (neutron star - neutron star) systems, and we know of 
no BH-NS (black hole - neutron star) nor BH-BH (black hole - black hole) system. The known NS-NS systems are listed along with their orbital parameters 
in Table~\ref{table:1}.

The observed NS-NS binaries have merger times $T_{merg}\succsim$~100~Myr.
However, a significant fraction of the population of the merging NS-NS 
may originate in the so-called ultra-compact NS-NS binaries, which have much shorter 
merger timescales $T_{merg}\precsim$ 100 Myr \citep{2002ApJ...571L.147B}. In addition, the small number 
of known pulsars are indicative of there being a significant fraction of very eccentric binaries.

Since no BH-NS nor BH-BH binaries are known, we can only rely on 
evolutionary considerations when estimating their number
and properties. It has been found that their number depends very strongly on the 
outcome of the common envelope phase when the secondary is on the Hertzsprung gap.
This phase will very likely end up as a merger and the formation of
a Thorne-Zytkow object. However, it has been demonstrated that in a low-metallicity environment the
common envelope mergers may be (to some extent) avoided and the BH-BH
formation is very effective \citep{2010ApJ...715L.138B}.

The eccentricity of a compact object binary may potentially be derived by analyzing
the inspiral signal, provided that the eccentricity is significant.
In this paper, we investigate the eccentricity distributions in the frequency band
of the currently working and future detectors of gravitational waves.
For the currently working detectors (LIGO and VIRGO), we assume 
that the sensitivity of the detectors will allow us to measure 
the signal for the frequencies starting at 30~Hz. This may not be 
accurate for the current state of these instruments, but it does 
accurately represent the predicted sensitivity of the Advanced LIGO/VIRGO detectors.
We consider two future detectors: the Einstein Telescope \citep{2010arXiv1003.1386V} and DECIGO \citep{2006AstHe..99..490K,2001PhRvL..87v1103S}.
For the Einstein Telescope, we assume that binaries shall be detectable 
from 3~Hz, and for DECIGO we assume that the lowest frequency detectable 
is 0.3~Hz. In all cases, these have to be treated as indicative numbers
that roughly describe these instruments. 

In section 2, we describe  the
model used to investigate the population of compact object binaries.
Section 3 presents the results for the current and future gravitational
wave detectors. In section 4, we summarize and discuss the results.

\begin{table*}
\caption{Known merging compact object binaries}
\label{table:1}
\centering
\begin{tabular}{c c c c c c c l}          
\hline\hline                        
Name & $P_{orb}~[h] $ & Present $e$ & $T_{merge}~[Gyr] $  & $e$ at 0.3~Hz  & $e$ at 3~Hz & $e$ at 30~Hz & Ref.\\   
\hline                                   
J0737-3039A/B & 2.454  & 0.088 & 0.085 & $4.5 \times 10^{-5}$ & $4 \times 10^{-6}$ & $3.5 \times 10^{-7}$ & \citet{2003Natur.426..531B}\\ 
B2127+11C & 8.05 & 0.681 & 0.2 & $2.9 \times 10^{-4}$ & $2.6 \times 10^{-5}$ & $2.3 \times 10^{-6}$ & \citet{1990Natur.346...42A}\\
J1906+0746 & 3.98 & 0.085 & 0.3 & $2.6 \times 10^{-5}$ & $2.3 \times 10^{-6}$ & $2 \times 10^{-7}$ & \citet{2006ApJ...640..428L}\\
B1913+16 & 7.752 & 0.617 & 0.3 & $2.2 \times 10^{-4}$ & $1.9 \times 10^{-5}$ & $1.7 \times 10^{-6}$ &\citet{2005ASPC..328...25W}\\
J1756-2251 & 7.67 & 0.181 & 1.7 & $2.6 \times 10^{-5}$ & $2.5 \times 10^{-6}$ & $2.2 \times 10^{-7}$ &\citet{2005ApJ...618L.119F}\\
B1534+12 (=J1537+1155) & 10.098 & 0.274 & 2.7 & $3.6 \times 10^{-5}$ & $3.2 \times 10^{-6}$ & $2.8 \times 10^{-7}$ &\citet{1991Natur.350..688W}\\
\hline                                             
\end{tabular}
\end{table*}

\section{The model}

\subsection{Compact object binary population model}
To model the population of compact object binaries,
we used the {\tt StarTrack} population synthesis code \citep{2002ApJ...572..407B}. It perform a suite
of Monte Carlo simulations of the stellar evolution of stars in environments
of two typical metallicities: $Z=Z_{\odot}=0.02$ and $Z=10\% \, Z_{\odot}=0.002$ 
\citep[e.g.,][]{2010ApJ...715L.138B}.
In these calculations, we employed the recent estimates of mass loss rates 
\citep{2010ApJ...714.1217B}. 
We calculate a population of 2 million massive binary stars, 
tracking the ensuing formation of relativistic
binary compact objects: double neutron stars (NS-NS), double black hole binaries 
(BH-BH), and mixed systems (BH-NS). Our modeling utilizes updated stellar and 
binary physics, including results from supernova simulations \citep{2001ApJ...554..548F}
 and compact object formation \citep{1996ApJ...457..834T}, incorporating elaborate 
mechanisms for treating stellar interactions such as mass transfer episodes \citep{2008ApJS..174..223B} 
or tidal synchronization and circularization \citep{1981A&A....99..126H}. We place
special emphasis on the common envelope evolution phase \citep{1984ApJ...277..355W}, which is
crucial for close double compact object formation because the attendant mass transfer 
permits an efficient hardening of the binary. This orbital contraction can be
sufficiently efficient to cause the individual stars in the binary to coalesce
and form a single highly rotating object, thereby preventing additional binary 
evolution and the formation of a double compact object. Because of 
significant radial expansion, stars crossing the Hertzsprung gap (HG) very 
frequently initiate a common envelope phase. 
HG stars do not have a clear 
entropy jump at the core-envelope transition \citep{2004ApJ...601.1058I}; if such a 
star overflows its Roche lobe and initiates a common envelope phase, the 
inspiral is expected to lead to a coalescence \citep{2000ARA&A..38..113T}. In 
particular, it has been estimated that for a solar metallicity environment (e.g., 
our Galaxy), properly accounting for the HG gap may lead to a reduction in the 
merger rates of BH-BH binaries by $\sim 2-3$ orders of magnitude \citep{2007ApJ...662..504B}.
In contrast, in a low metallicity environment this suppression is much less
severe ($\sim 1$ order of magnitude; \citet{2010ApJ...715L.138B}).
The details of the common envelope phase are not yet fully understood, 
thus in what follows we consider two set of models, one that does not take into 
account the suppression (optimistic models: marked with A), and another that assumes 
the maximum suppression (pessimistic models: marked with B). Solar metallicity and
$10\%$ of solar metallicity are labeled with Z and z, respectively.
In the case of NSs, we adopt natal kick distributions from observations of single 
Galactic pulsars \citep{2005MNRAS.360..974H} with $\sigma =265$~km/s. 
However, for BHs we draw kicks from the same distribution (but at a lower
magnitude), which is inverse proportional to the amount of fall back expected at BH 
formation \citep[e.g.,][]{2001ApJ...554..548F}. In particular, for most massive BHs
that form with the full fall back (direct BH formation), the amount of
natal kick is zero.
In addition, we test one more set of models in which the magnitude of the NS kicks
is lower by a factor of 2, to $\sigma=132.5$~km/s, as some observations and empirically 
based arguments seem to indicate that natal kicks in close binaries are
lower than for single stars \citep{2006ApJ...644.1063D,2006A&A...450..345K}. 
The BH kicks are decreased in the similar fashion as in models with the full NS kicks.
The standard value of $\sigma$ parameter is denoted by K and the smaller
value by k.
The detailed list of models considered in this paper is presented in Table~\ref{Models}.
Model AZK is a standard set of parameters described in detail by \citet{2002ApJ...572..407B}.

\begin{table}[ht]
\caption{The list of models of stellar evolution used in the paper.}
\label{Models}
\begin{minipage}{\linewidth}\centering
\begin{tabular}{c c c c}
\hline\hline
Model & Metallicity & $\sigma$ $[kms^{-1}]$ & HG \\
\hline
AZK & $Z_{\odot}$ & $265.0$ & + \\
BZK & $Z_{\odot}$ & $265.0$ & - \\
AZk & $Z_{\odot}$ & $132.5$ & + \\
BZk & $Z_{\odot}$ & $132.5$ & - \\
AzK & $10\%$ $Z_{\odot}$ & $265.0$ & + \\
BzK & $10\%$ $Z_{\odot}$ & $265.0$ & - \\
Azk & $10\%$ $Z_{\odot}$ & $132.5$ & + \\
Bzk & $10\%$ $Z_{\odot}$ & $132.5$ & - \\
\hline
\end{tabular}
\end{minipage}
\end{table}

\subsection{Evolution of orbits}

The evolution
of the orbit of compact object binary under the influence of gravitational radiation
had been calculated by \citet{1963PhRv..131..435P,1964PhRv..136.1224P}. In the quadrupole approximation,
the orbit decays as

\begin{eqnarray}
\label{dadt}
\frac{da}{dt}=- \frac{\beta}{a^3} \Psi(e), \hspace{1cm}
\Psi(e)=\frac{1+73/24 e^2 + 37/96 e^4}{(1-e^2)^{7/2}},
\end{eqnarray}
where $a$ is the great semi-axis, $e$ is the eccentricity of binary, $M_1$ is the mass of the first component,
$M_2$ is the mass of second component, and 

\begin{eqnarray}
\beta=\frac{64}{5}\frac{G^3 \mu M^2}{c^5}, \hspace{1cm} \mu=\frac{M_1 M_2}{M_1+M_2}.
\end{eqnarray}
While the eccentricity decays as

\begin{eqnarray}
\label{dedt}
\frac{de}{dt}=- \frac{19}{12}\frac{\beta}{a^4} \Phi(e), \hspace{1cm}
\Phi(e)=\frac{(1+121/304 e^2)e}{(1-e^2)^{5/2}}.
\end{eqnarray}

Using the above formulae we can express the fundamental gravitational wave frequency  
as a function of the eccentricity
\begin{eqnarray}
\label{f(e)}
f_{GW}(e)=\frac{2}{P_0} \frac{(1-e^2)^{3/2}}{e^{18/19}}[1+\frac{121}{304}e^2]^{-1305/2299} c_0^{3/2},
\end{eqnarray}
where $c_0=(e_0^{12/19}[1+\frac{121}{304}e_0^2]^{1305/2299})(1-e_0^2)^{-1}$, $P_0$ is the initial orbital period,
and $f_{GW}(e)$ is the first non-zero harmonic.
The gravitational wave frequency is twice the orbital frequency, i.e., $f_{GW}=2f_{orb}=\frac{2}{P_{orb}}$.

\begin{figure}
\includegraphics[scale=0.7,angle=270]{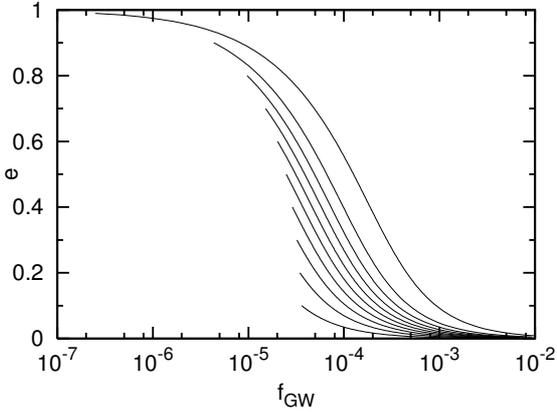}
\caption{We present ten cases of eccentricity evolution, starting with different values of e from $e=0.1$ (first line from the bottom) to $e=0.99$ (first line from the top). Initial semi-major axis is chosen such that a binary will merge within time $T_{merg}=10\, Gyr$ in each case.
}
\label{Ecevol}
\end{figure}

We present the evolution of eccentricity as a function of gravitational wave frequency in
Figure~\ref{Ecevol} for a binary neutron star with components of equal masses of $ 1.4 \,M_{\odot}$. 
The initial frequency corresponds to a semi-major axis such that the merger time is set to be $T_{merg}=10^4$~Myr.
Figure~\ref{Ecevol} contains several different cases of evolution in the plane stretched by eccentricity and gravitational wave frequency.

\section{Results}

\begin{figure*}
\centering
\includegraphics[scale=0.75, angle=270]{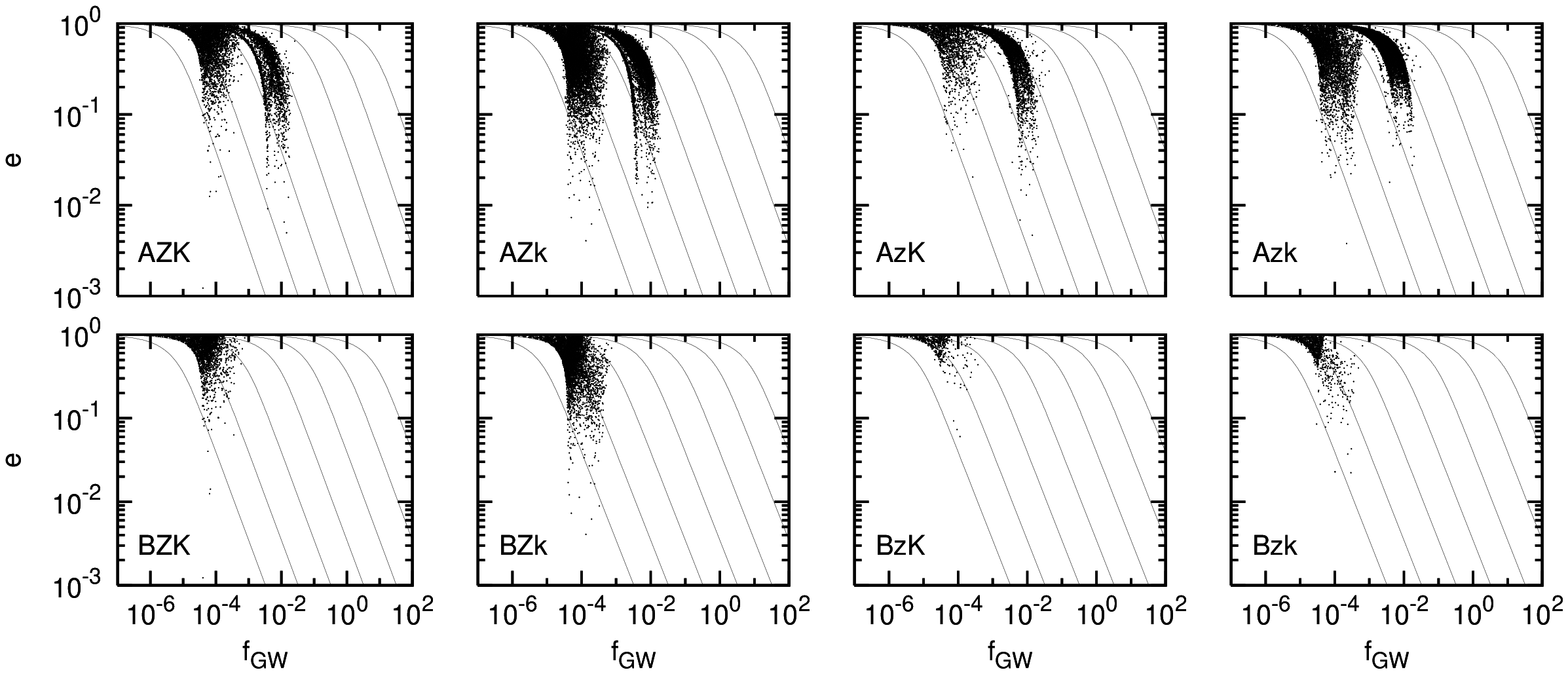}
\caption{The properties of the population of double neutron stars 
obtained using the {\tt StarTrack} code. The plot shows only the binaries 
that will merge within the Hubble time. Solid lines correspond to evolutionary
tracks for initial gravitational waves frequencies from $f_{0}=10^{-8}$~Hz (first line from the left-hand side) 
to $f_0=10^2$~Hz (first line from the right-hand side).}
\label{Initial_nsns}
\end{figure*}

\begin{figure*}
\centering
\includegraphics[scale=0.9, angle=270]{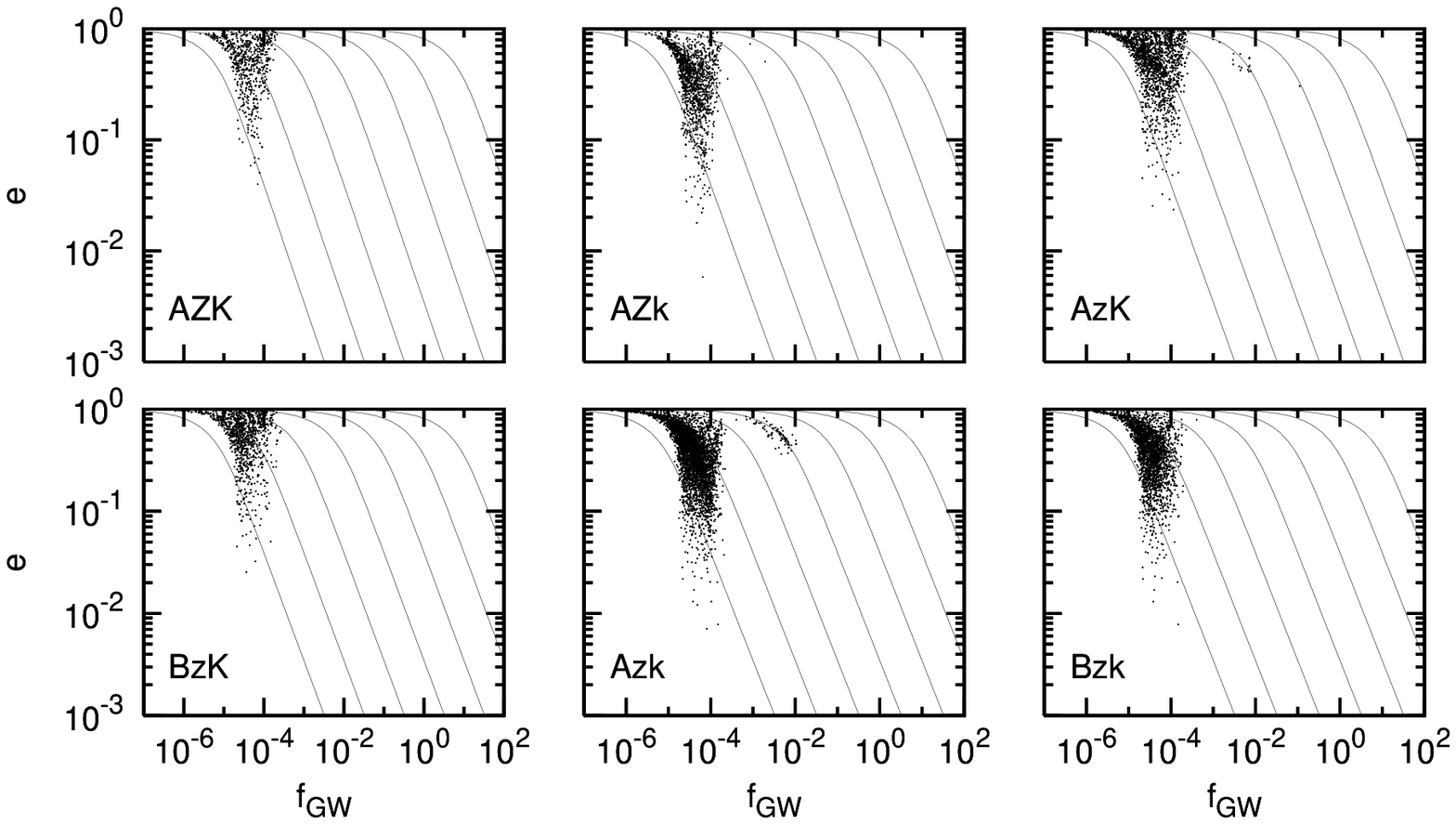}
\caption{The properties of the population of neutron star - black hole systems
obtained using the {\tt StarTrack} code. The plot shows only the binaries 
that will merge within the Hubble time. Solid lines correspond to evolutionary
tracks for initial gravitational wave frequencies from $f_{0}=10^{-8}$~Hz (first line from the left-hand side) 
to $f_0=10^2$~Hz (first line from the right-hand side).}
\label{Initial_nsbh}
\end{figure*}

\begin{figure*}
\centering
\includegraphics[scale=0.9, angle=270]{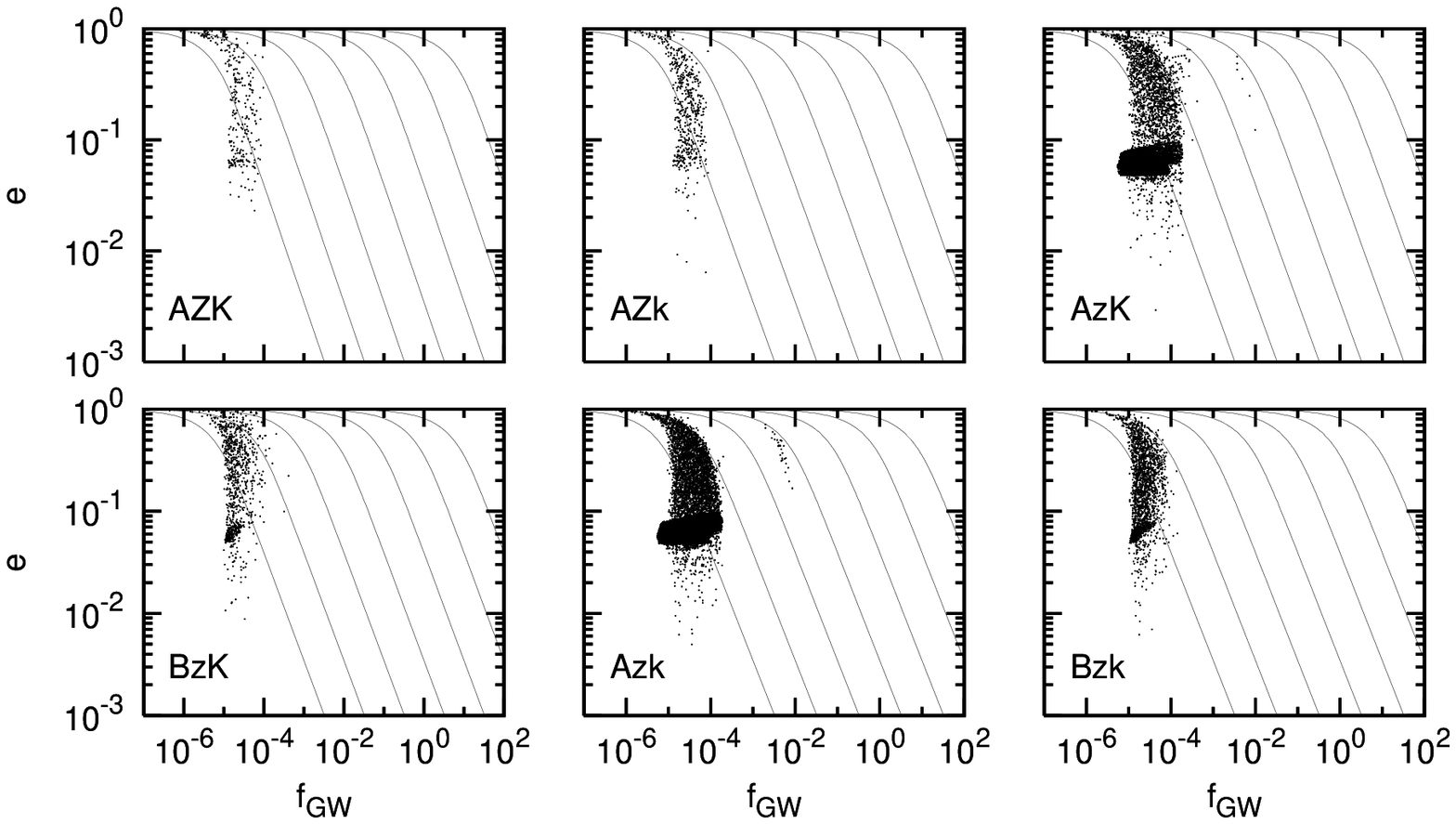}
\caption{The properties of the population of double black holes 
obtained using the {\tt StarTrack} code. The plot shows only the binaries 
that will merge within the Hubble time. Solid lines correspond to evolutionary
tracks for initial gravitational waves frequencies from $f_{0}=10^{-8}$~Hz (first line from the left-hand side)
 to $f_0=10^2$~Hz (first line from the right-hand side).}
\label{Initial_bhbh}
\end{figure*}

\subsection{Properties of the binaries at formation time}

We start with an initial population created using the {\tt StarTrack} code.
We present the properties of the population of compact object binaries 
in Figures~\ref{Initial_nsns} - \ref{Initial_bhbh} in the space spanned by the initial eccentricity
and initial gravitational wave frequency, which is twice the orbital frequency.
Each panel in theses figures corresponds to a different model labeled 
as listed in Table~\ref{Models}.

The case of the NS-NS systems is shown in Figure~\ref{Initial_nsns}.
The boundary of the region populated by the systems on the left-hand side corresponds to 
the requirement that we only consider binaries that merge within a Hubble time.
The bulk of the binaries shown in each panel correspond to those
that have undergone one CE phase in their evolution.
The top row corresponds to the models AZK, AZk, AzK, and Azk, in which 
we allow the binaries to cross through the common envelope with
the donor on the Hertzsprung gap, denoted by "+" in Table~\ref{Models}.
These binaries may undergo a second common envelope  phase with 
a helium star companion. At the second CE stage, the orbit is tightened even more
leading to formation of the stripe in the diagram stretching from 
$f_{GW}\approx 10^{-2}$~Hz at $e\approx 10^{-2}$. 
In these models, the initial distribution in the 
space of gravitational wave frequency versus eccentricity is bimodal.
The influence of the value of the kick velocity has a small impact on the 
shape of distributions presented in Figure~\ref{Initial_nsns} as can be seen 
by comparing the data in plots labeled as either K-large kicks or k-small kicks.

For BH-NS systems, presented in Figure~\ref{Initial_nsbh}, and BH-BH binaries,
in Figure~\ref{Initial_bhbh}, we present the results of six out of eight 
models, since in models BZK and BZk, almost no binaries are formed in our simulations
that involve $2\times 10^6$ initial binaries.
For BH-NS and BH-BH binaries, the formation of ultra-compact binaries is not
expected. The formation of NS-NS ultra-compact systems in very close orbits is the
consequence of the final CE episode, which is initiated by a low-mass helium (2-4
$M_{\odot}$) star and its NS companion ($1.4\, M_{\odot}$). Since the donor is about twice
as massive as its companion, the CE phase is initiated by the non-stable mass transfer and the orbit
significantly decreased in size.
For more massive BH-BH/BH-NS binaries, helium stars are on average
more massive ($M>3-4\, M_{\odot}$) and do not expand (so no CE phase), and even if 
a low mass helium star forms, then its companion is a BH ($M>3\, M_{\odot}$), so most likely instead of
CE the RLOF is stable and does not lead to orbital decay (mass ratio
$\sim 1$). Very few systems (e.g., models AzK or Azk) produce ultra-compact 
BH-NS/BH-BH binaries for very special cases of binary evolution.
In the case of BH-BH binaries, shown in Figure~\ref{Initial_bhbh} we present only 
six models, since models BZK and BZk do not lead to the formation of BH-BH binaries
\citep{2007ApJ...662..504B}.
In all models, there is an enhanced density of systems 
formed with $e\approx 0.1$ at approximately
$10^{-5}\,{\rm Hz} < f_{GW} < 10^{-4}\,{\rm Hz}$. In these systems, 
the second black hole has formed via direct collapse. When treating the 
direct collapse, we assume that 10\% of the mass escapes 
 in the form of neutrinos and possibly gravitational waves. Hence, the 
gravitational mass of the BH is 10\% lower than the baryon mass of the collapsing star.
This introduces a small eccentricity $\approx 0.1$ 
since the systems were circularized in the mass transfer prior to the collapse
and the formation of the second BH.

\subsection{Eccentricity when binary enters detector band}

\begin{figure}

\subfloat{\includegraphics[scale=0.7, angle=270]{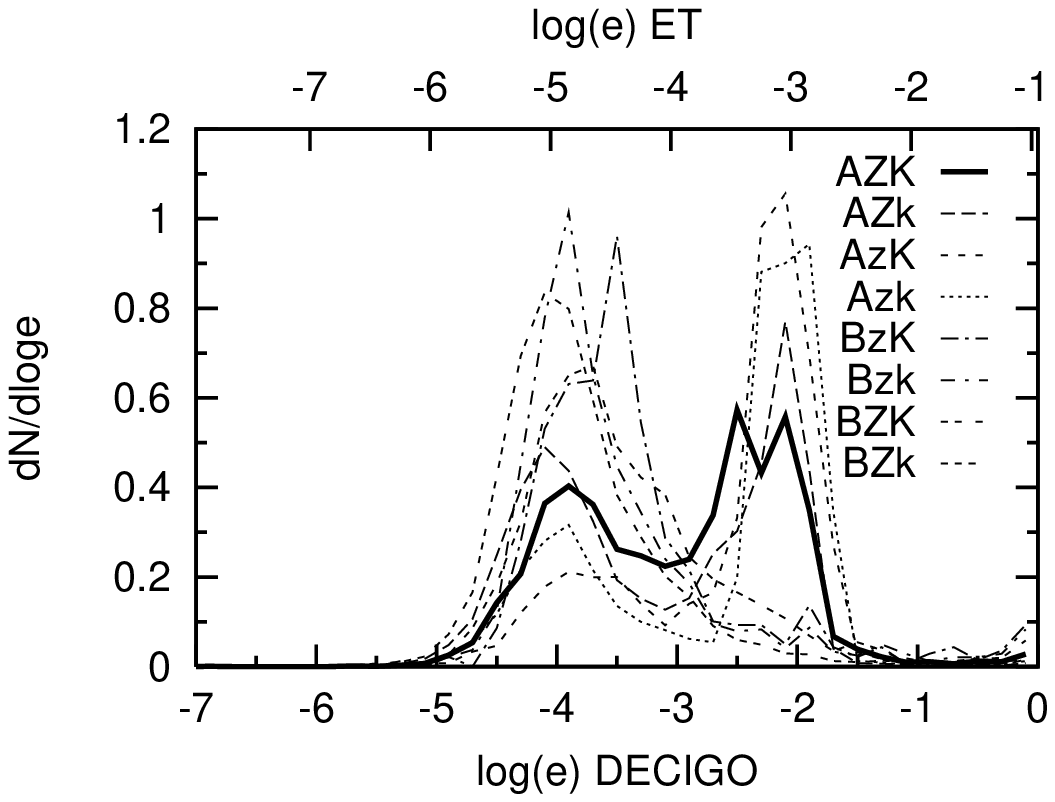}}\\
\subfloat{\includegraphics[scale=0.7, angle=270]{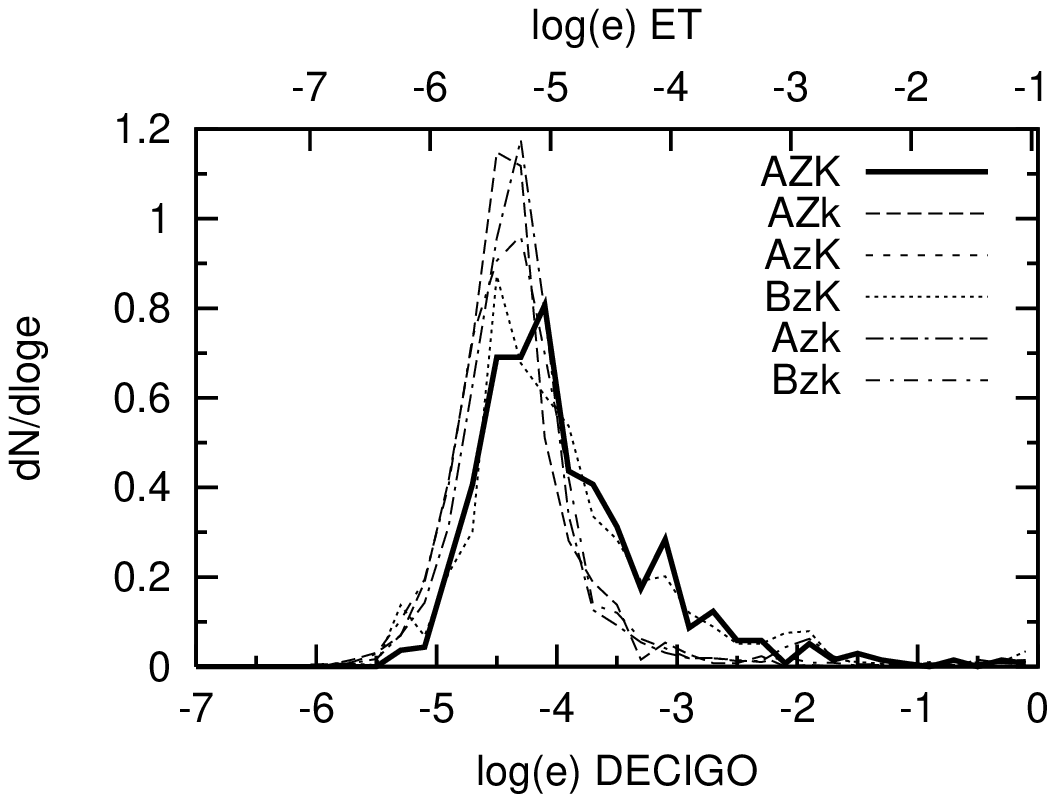}}\\
\subfloat{\includegraphics[scale=0.7, angle=270]{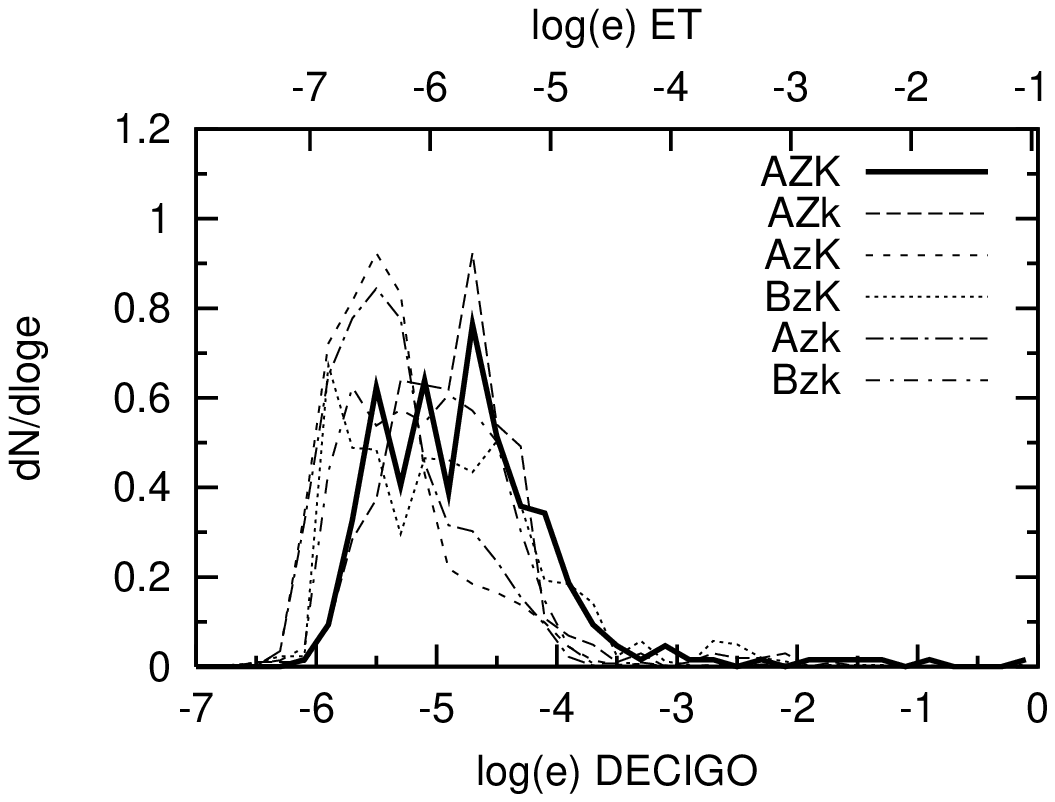}}
\caption{Distribution of eccentricity for (NS-NS - top panel, BH-NS middle panel and BH-BH bottom panel) seen at 0.3~Hz (DECIGO -like detectors) and at 3~Hz (ET - like detectors). Solid thick line corresponds to standard model (AZK). Dashed and dotted lines indicate other models.}
\label{Results}
\end{figure}

For the detection of gravitational waves, 
it is important to know the eccentricity of a binary at the time it enters the 
sensitivity window of the detector.
We consider three cases that correspond approximately to 
three types of detectors. 
In Figure~\ref{Results}, we show the eccentricity distributions 
at 0.3~Hz (bottom horizontal axis) and 3~Hz (top horizontal axis) corresponding approximately to 
the ET and DECIGO detectors.
The results for 
the Advanced LIGO/VIRGO can be easily obtained by rescaling the horizontal axis.

The shape of the eccentricity distributions at the moment that the binary enters the given
detector band follows
from the corresponding initial distribution.
However, one must note that for each type of binary 
there is a different natural timescale and frequency,
because of the different mass scales of each binary.

We present the results for the DECIGO detector and 
add appropriate numbers for the ET in parentheses.
For the NS-NS binaries shown in the top panel 
of Figure~\ref{Results}, the distribution is 
either centered on  $e\approx 10^{-4}$ (ET: $10^{-5}$)
 for the models
BZK, BZk, BzK, and Bzk, where we do not allow the formation of ultra-compact 
binaries in a second CE phase. The remaining models AZK, AZk, AzK, and Azk 
contain another component centered roughly at $e\approx 10^{-4}$ (ET: $10^{-3}$). This additional component represent
the ultra-compact binaries that have experienced two episodes 
of mass transfer in their evolutionary history and were already very tight at the second supernova explosion.
The mixed BH-NS binaries, shown in 
the middle panel of Figure~\ref{Results} exhibit  
a distribution of eccentricity centered at $e\approx 3\times 10^{-5}$ (ET: $3\times10^{-6}$), 
while the eccentricity BH-BH binaries, shown in the 
bottom panel of Figure~\ref{Results} lie between 
$e\approx \times 10^{-6}$(ET: $10^{-7}$) and $e\approx \times 10^{-4}$ (ET: $10^{-5}$) .

For the Advanced LIGO/VIRGO detectors where we assume that the low 
frequency boundary lies at $\approx$ 30 Hz, the eccentricities are even smaller.
It follows from equation \ref{f(e)} that the distributions are shifted 
by a factor of $10^{-19/18}$ for each factor of ten in 
frequency. Thus, the values of eccentricity in the case of Advanced LIGO/VIRGO type detectors
 are consistent with $e=0$ and 
we can safely assume that all BH-NS and BH-BH binaries 
are circular without any loss of sensitivity.

In table~\ref{fractions}, we present the fraction of binaries 
with eccentricities above 0.01 at the time of entering the detector band,
to help quantify the extent of the large eccentricity tails of the distributions presented in Figure~\ref{Results}.
This fraction does not reflect the detectability of eccentricity \citep{2010arXiv1006.3759S},which for realistic 
distributions of binaries will be discussed in a forthcoming paper.

\begin{table}
\caption{Fraction 
 of the compact binaries with eccentricity greater than $10^{-2}$. 
Top table corresponds to double neutron stars (NS-NS), middle to the mixed systems (BH-NS),
and the bottom to the binary black holes (BH-BH). 
We present the fraction  of binaries at the moment 
of  entering  different frequency bands (30 Hz, 3 Hz, and 0.3 Hz). 
In brackets, we include the number of these  systems in the simulation N. 
We only listed results for models that are non-zero.
The number of digits shown is for formatting only, and the relative sampling error is $N^{-1/2}$.}
\label{results_table}
\begin{minipage}{\linewidth}\centering
\begin{tabular}{c r@{} @{}l r r@{} @{}l r r@{} @{}l r}
\hline\hline
\multicolumn{10}{c}{NS-NS} \\
\hline\hline
 & \multicolumn{3}{c}{30 Hz} & \multicolumn{3}{c}{3 Hz} & \multicolumn{3}{c}{0.3 Hz} \\
\hline
AZK & 0.&60\%& (51) & 1.&32\%& (112) & 11.&13\%& (945) \\
BZK & 1.&27\%& (36) & 2.&33\%& (66) & 6.&52\%& (185) \\
AZk & 0.&16\%& (27) & 0.&38\%& (64) & 10.&37\%& (1732) \\
BZk & 0.&30\%& (15) & 0.&75\%& (37) & 2.&22\%& (110) \\
AzK & 0.&29\%& (25) & 0.&96\%& (83) & 21.&74\%& (1880) \\
BzK & 1.&87\%& (13) & 4.&02\%& (28) & 9.&33\%& (65) \\
Azk & 0.&26\%& (37) & 0.&57\%& (81) & 26.&91\%& (3799) \\
Bzk & 1.&74\%& (21) & 3.&31\%& (40) & 7.&79\%& (94) \\
\end{tabular}
\end{minipage}

\begin{minipage}{\linewidth}\centering
\begin{tabular}{c r@{} @{}l r r@{} @{}l r r@{} @{}l r}
\hline\hline
\multicolumn{10}{c}{BH-NS} \\
\hline\hline
& \multicolumn{3}{c}{30 Hz} & \multicolumn{3}{c}{3 Hz} & \multicolumn{3}{c}{0.3 Hz} \\
\hline
AZK & 0.&29\%& (2) & 0.&73\%& (5) & 3.&05\%& (21) \\
AZk & 0.&15\%& (2) & 0.&54\%& (7) & 0.&61\%& (8) \\
AzK & 0.&56\%& (14) & 0.&96\%& (24) & 3.&96\%& (99) \\
BzK & 0.&68\%& (10) & 1.&23\%& (18) & 3.&63\%& (53) \\
Azk & 0.&35\%& (15) & 0.&78\%& (34) & 2.&81\%& (122) \\
Bzk & 0.&33\%& (8) & 0.&91\%& (22) & 1.&53\%& (37) \\
\end{tabular}
\end{minipage}

\begin{minipage}{\linewidth}\centering
\begin{tabular}{c r@{} @{}l r r@{} @{}l r r@{} @{}l r}
\hline\hline
\multicolumn{10}{c}{BH-BH} \\
\hline\hline
& \multicolumn{3}{c}{30 Hz} & \multicolumn{3}{c}{3 Hz} & \multicolumn{3}{c}{0.3 Hz} \\
\hline
AZK & 0.&31\%& (1) & 0.&62\%& (2) & 1.&87\%& (6) \\
AzK & 0.&02\%& (3) & 0.&02\%& (4) & 0.&13\%& (23) \\
BzK & 0.&15\%& (2) & 0.&15\%& (2) & 0.&46\%& (6) \\
Azk & 0.&00\%& (1) & 0.&01\%& (2) & 0.&03\%& (6) \\
\hline
\end{tabular}
\end{minipage}
\label{fractions}
\end{table}

\section{Summary}
We have presented the eccentricity distributions of
compact object binaries at three frequencies immediately before merger.
The properties of the compact object binaries have been calculated using the 
{\tt StarTrack} population synthesis code. We have found that the eccentricity distributions of the compact object binaries do not depend strongly
on the assumed model of binary evolution.
Any dependence has been found to be the strongest for binary neutron stars, whose
distributions may be either
single or double peaked. The extra peak corresponds
to ultra-compact NS-NS binaries that have undergone an additional CE 
phase immediately before forming the second NS.

To make the results easier to use in the simulations,
we have fitted the resulting distributions of eccentricity with 
 a single log-normal distribution in the case of BH-BH and BH-NS binaries
\begin{eqnarray}
\label{fit}
f(x)=\frac{1}{\sigma \sqrt{2\pi}} \exp\left(-\frac{(x-\mu)^2}{2\sigma^2}\right),
\end{eqnarray}
where $x=\log{e}$, $\mu$ is the mean, and $\sigma$ is the variance.

For
the NS-NS eccentricities, we used a sum of two log-normal distributions with two weights,
since the distribution is double peaked, given by
\begin{eqnarray}
\label{fit2}
f(x)=\frac{w}{\sigma_1 \sqrt{2\pi}} \exp\left(-\frac{(x-\mu_1)^2}{2\sigma_1^2}\right) + \frac{(1-w)}{\sigma_2 \sqrt{2\pi}} \exp\left(-\frac{(x-\mu_2)^2}{2\sigma_2^2}\right),
\end{eqnarray}
where $x=\log{e}$, $\mu_1$ is the mean of the first peak, $\mu_2$ is the mean of the second peak, $\sigma_1$ is the variance of the first distribution,
$\sigma_2$ is the variance of the second distribution, and $w$ is the weight.

We used the Marquardt-Levenberg algorithm to find the parameters, and estimate the asymptotic standard error of each of them.
The results of the fits are shown in Table~\ref{fit_table}. The widths of the distributions $\sigma$
are the same for each frequency band and only the centroids move.
\begin{table}[ht]
\caption{Parameters of log normal distribution fitted to results of model AZK with asymptotic standard errors.}
\label{fit_table}
\begin{minipage}{\linewidth}\centering
\begin{tabular}{c c c c}
\hline\hline
\multicolumn{4}{c}{NS-NS} \\
\hline\hline

 & $\sigma_1$: & 0.47 $\pm$ 0.04 & \\
 &  $\sigma_2$: & 0.39 $\pm$ 0.03 &\\
& $w$: & 0.46 $\pm$ 0.03 &\\
\hline
 & 0.3 Hz & 3 Hz & 30 Hz \\
\hline
$\mu_1$ & -3.83 $\pm$ 0.04 & -4.89 $\pm$ 0.04 & -5.94 $\pm$ 0.04\\
$\mu_2$ & -2.32 $\pm$ 0.03 & -3.38 $\pm$ 0.03 & -4.43 $\pm$ 0.03\\
\end{tabular}
\end{minipage}
\begin{minipage}{\linewidth}\centering
\begin{tabular}{c c c c}
\hline\hline
\multicolumn{4}{c}{BH-NS} \\
\hline\hline
 & $\sigma$: & 0.55 $\pm$ 0.03 & \\
\hline
 & 0.3 Hz & 3 Hz & 30 Hz \\
\hline
$\mu$ & -4.18 $\pm$ 0.04 & -5.23 $\pm$ 0.03 & -6.27 $\pm$ 0.03\\
\end{tabular}
\end{minipage}
\begin{minipage}{\linewidth}\centering
\begin{tabular}{c c c c}
\hline\hline
\multicolumn{4}{c}{BH-BH} \\
\hline\hline
 &  $\sigma$: & 0.70 $\pm$ 0.03 & \\
\hline
& 0.3 Hz & 3 Hz & 30 Hz \\
\hline
$\mu$ & -4.91 $\pm$ 0.08 & -5.95 $\pm$ 0.03 & -7.06 $\pm$ 0.03\\
\hline
\end{tabular}
\end{minipage}
\end{table}

The eccentricity of the BH-BH binaries in all three 
cases is negligible. This is due to two factors.
First, at  the frequencies of interest the BH-BH systems 
are much closer to coalescence than the neutron star.
Second, the initial kicks at formation of BHs are lower 
than in the case of NS, so the initial eccentricities of 
BH-BH systems are typically lower than in the case of NS-NS ones.

The eccentricities of the mixed BH-NS systems are larger
than in the case of BH-BH ones. However, the number of systems 
is small enough to ensure that 
by neglecting the eccentricity we do not decrease the sensitivity of 
Advanced LIGO/VIRGO detectors. 
For the ET-like detector,
some eccentric systems may be detected. In the case of 
the DECIGO-like detector, the number of  systems with eccentricities 
above 0.01 lies between 3\% and 4\%.

The eccentricity of NS-NS systems are larger than those of binaries containing BHs.
Given a much larger expected detection rate for ET, this 
means that there should be a significant number of NS-NS binaries with 
detectable eccentricities. Finally, in the case of 
DECIGO a fraction of between 2\% and 27\%
of the NS-NS binaries have eccentricities above 0.01.
Moreover, the shape of the eccentricity distribution 
of NS-NS binaries will depend on the existence of an evolutionary scenario 
leading to the formation of ultra-compact binaries.
Thus, the measurement of the eccentricity distribution is an interesting tool
for probing the details of NS-NS formation scenarios.

\section*{Acknowledgments}
 This work was supported by the
EGO-DIR-102-2007; the FOCUS 4/2007 Program of Foundation for Polish
Science, the Polish grants N N203 511238, DPN/N176/VIRGO/2009,
N N203 302835, N N203 404939 and by CompStar a Research Networking Programme of
the European Science Foundation.

\bibliography{15777}

\begin{thebibliography}{44}
\expandafter\ifx\csname natexlab\endcsname\relax\def\natexlab#1{#1}\fi

\bibitem[{{Abadie} {et~al.}(2010){Abadie}, {Abbott}, {Abbott}, {Abernathy},
  {Accadia}, {Acernese}, {Adams}, {Adhikari}, {Ajith}, {Allen}, \&
  et~al.}]{2010CQGra..27q3001A}
{Abadie}, J., {Abbott}, B.~P., {Abbott}, R., {et~al.} 2010, Classical and
  Quantum Gravity, 27, 173001

\bibitem[{{Acernese} {et~al.}(2006){Acernese}, {Amico}, {Al-Shourbagy},
  {Aoudia}, {Avino}, {Babusci}, {Ballardin}, {Barone}, {Barsotti}, {Barsuglia},
  {Beauville}, {Bizouard}, {Boccara}, {Bondu}, {Bosi}, {Bradaschia},
  {Birindelli}, {Braccini}, {Brillet}, {Brisson}, {Brocco}, {Buskulic},
  {Calloni}, {Campagna}, {Cavalier}, {Cavalieri}, {Cella}, {Chassande-Mottin},
  {Corda}, {Clapson}, {Cleva}, {Coulon}, {Cuoco}, {Dattilo}, {Davier}, {De
  Rosa}, {Di Fiore}, {Di Virgilio}, {Dujardin}, {Eleuteri}, {Enard},
  {Ferrante}, {Fidecaro}, {Fiori}, {Flaminio}, {Fournier}, {Francois},
  {Frasca}, {Frasconi}, {Freise}, {Gammaitoni}, {Gennai}, {Giazotto},
  {Giordano}, {Giordano}, {Gouaty}, {Grosjean}, {Guidi}, {Hebri}, {Heitmann},
  {Hello}, {Holloway}, {Karkar}, {Kreckelbergh}, {La Penna}, {Letendre},
  {Lorenzini}, {Loriette}, {Loupias}, {Losurdo}, {Mackowski}, {Majorana},
  {Man}, {Mantovani}, {Marchesoni}, {Marion}, {Marque}, {Martelli}, {Masserot},
  {Mazzoni}, {Milano}, {Moins}, {Moreau}, {Morgado}, {Mours}, {Pai}, {Palomba},
  {Paoletti}, {Pardi}, {Pasqualetti}, {Passaquieti}, {Passuello}, {Perniola},
  {Piergiovanni}, {Pinard}, {Poggiani}, {Punturo}, {Puppo}, {Qipiani},
  {Rapagnani}, {Reita}, {Remillieux}, {Ricci}, {Ricciardi}, {Ruggi}, {Russo},
  {Solimeno}, {Spallicci}, {Stanga}, {Taddei}, {Tonelli}, {Toncelli},
  {Tournefier}, {Travasso}, {Vajente}, {Verkindt}, {Vetrano}, {Vicer{\'e}},
  {Vinet}, {Vocca}, {Yvert}, \& {Zhang}}]{2006CQGra..23S..63A}
{Acernese}, F., {Amico}, P., {Al-Shourbagy}, M., {et~al.} 2006, Classical and
  Quantum Gravity, 23, 63

\bibitem[{{Anderson} {et~al.}(1990){Anderson}, {Gorham}, {Kulkarni}, {Prince},
  \& {Wolszczan}}]{1990Natur.346...42A}
{Anderson}, S.~B., {Gorham}, P.~W., {Kulkarni}, S.~R., {Prince}, T.~A., \&
  {Wolszczan}, A. 1990, \nat, 346, 42

\bibitem[{{Belczynski} {et~al.}(2010{\natexlab{a}}){Belczynski}, {Bulik},
  {Fryer}, {Ruiter}, {Valsecchi}, {Vink}, \& {Hurley}}]{2010ApJ...714.1217B}
{Belczynski}, K., {Bulik}, T., {Fryer}, C.~L., {et~al.} 2010{\natexlab{a}},
  \apj, 714, 1217

\bibitem[{{Belczynski} {et~al.}(2002{\natexlab{a}}){Belczynski}, {Bulik}, \&
  {Kalogera}}]{2002ApJ...571L.147B}
{Belczynski}, K., {Bulik}, T., \& {Kalogera}, V. 2002{\natexlab{a}}, \apjl,
  571, L147

\bibitem[{{Belczynski} {et~al.}(2010{\natexlab{b}}){Belczynski}, {Dominik},
  {Bulik}, {O'Shaughnessy}, {Fryer}, \& {Holz}}]{2010ApJ...715L.138B}
{Belczynski}, K., {Dominik}, M., {Bulik}, T., {et~al.} 2010{\natexlab{b}},
  \apjl, 715, L138

\bibitem[{{Belczynski} {et~al.}(2002{\natexlab{b}}){Belczynski}, {Kalogera}, \&
  {Bulik}}]{2002ApJ...572..407B}
{Belczynski}, K., {Kalogera}, V., \& {Bulik}, T. 2002{\natexlab{b}}, \apj, 572,
  407

\bibitem[{{Belczynski} {et~al.}(2008){Belczynski}, {Kalogera}, {Rasio}, {Taam},
  {Zezas}, {Bulik}, {Maccarone}, \& {Ivanova}}]{2008ApJS..174..223B}
{Belczynski}, K., {Kalogera}, V., {Rasio}, F.~A., {et~al.} 2008, \apjs, 174,
  223

\bibitem[{{Belczynski} {et~al.}(2007){Belczynski}, {Taam}, {Kalogera}, {Rasio},
  \& {Bulik}}]{2007ApJ...662..504B}
{Belczynski}, K., {Taam}, R.~E., {Kalogera}, V., {Rasio}, F.~A., \& {Bulik}, T.
  2007, \apj, 662, 504

\bibitem[{{Bogomazov} {et~al.}(2007){Bogomazov}, {Lipunov}, \&
  {Tutukov}}]{2007ARep...51..308B}
{Bogomazov}, A.~I., {Lipunov}, V.~M., \& {Tutukov}, A.~V. 2007, Astronomy
  Reports, 51, 308

\bibitem[{{Burgay} {et~al.}(2003){Burgay}, {D'Amico}, {Possenti}, {Manchester},
  {Lyne}, {Joshi}, {McLaughlin}, {Kramer}, {Sarkissian}, {Camilo}, {Kalogera},
  {Kim}, \& {Lorimer}}]{2003Natur.426..531B}
{Burgay}, M., {D'Amico}, N., {Possenti}, A., {et~al.} 2003, \nat, 426, 531

\bibitem[{{De Donder} \& {Vanbeveren}(2004)}]{2004NewA....9....1D}
{De Donder}, E. \& {Vanbeveren}, D. 2004, \na, 9, 1

\bibitem[{{Dessart} {et~al.}(2006){Dessart}, {Burrows}, {Ott}, {Livne}, {Yoon},
  \& {Langer}}]{2006ApJ...644.1063D}
{Dessart}, L., {Burrows}, A., {Ott}, C.~D., {et~al.} 2006, \apj, 644, 1063

\bibitem[{{Dewi} {et~al.}(2005){Dewi}, {Podsiadlowski}, \&
  {Pols}}]{2005MNRAS.363L..71D}
{Dewi}, J.~D.~M., {Podsiadlowski}, P., \& {Pols}, O.~R. 2005, \mnras, 363, L71

\bibitem[{{Dewi} {et~al.}(2002){Dewi}, {Pols}, {Savonije}, \& {van den
  Heuvel}}]{2002MNRAS.331.1027D}
{Dewi}, J.~D.~M., {Pols}, O.~R., {Savonije}, G.~J., \& {van den Heuvel},
  E.~P.~J. 2002, \mnras, 331, 1027

\bibitem[{{Faulkner} {et~al.}(2005){Faulkner}, {Kramer}, {Lyne}, {Manchester},
  {McLaughlin}, {Stairs}, {Hobbs}, {Possenti}, {Lorimer}, {D'Amico}, {Camilo},
  \& {Burgay}}]{2005ApJ...618L.119F}
{Faulkner}, A.~J., {Kramer}, M., {Lyne}, A.~G., {et~al.} 2005, \apjl, 618, L119

\bibitem[{{Fryer} \& {Kalogera}(2001)}]{2001ApJ...554..548F}
{Fryer}, C.~L. \& {Kalogera}, V. 2001, \apj, 554, 548

\bibitem[{{Gondek-Rosi{\'n}ska} {et~al.}(2007){Gondek-Rosi{\'n}ska}, {Bulik},
  \& {Belczy{\'n}ski}}]{2007AdSpR..39..285G}
{Gondek-Rosi{\'n}ska}, D., {Bulik}, T., \& {Belczy{\'n}ski}, K. 2007, Advances
  in Space Research, 39, 285

\bibitem[{{Harry} \& {the LIGO Scientific
  Collaboration}(2010)}]{2010CQGra..27h4006H}
{Harry}, G.~M. \& {the LIGO Scientific Collaboration}. 2010, Classical and
  Quantum Gravity, 27, 084006

\bibitem[{{Hobbs} {et~al.}(2005){Hobbs}, {Lorimer}, {Lyne}, \&
  {Kramer}}]{2005MNRAS.360..974H}
{Hobbs}, G., {Lorimer}, D.~R., {Lyne}, A.~G., \& {Kramer}, M. 2005, \mnras,
  360, 974

\bibitem[{{Hut}(1981)}]{1981A&A....99..126H}
{Hut}, P. 1981, \aap, 99, 126

\bibitem[{{Ivanova} \& {Taam}(2004)}]{2004ApJ...601.1058I}
{Ivanova}, N. \& {Taam}, R.~E. 2004, \apj, 601, 1058

\bibitem[{{Kawamura}(2006)}]{2006AstHe..99..490K}
{Kawamura}, S. 2006, Astronomical Herald, 99, 490

\bibitem[{{Kiel} {et~al.}(2010){Kiel}, {Hurley}, \&
  {Bailes}}]{2010MNRAS.406..656K}
{Kiel}, P.~D., {Hurley}, J.~R., \& {Bailes}, M. 2010, \mnras, 406, 656

\bibitem[{{Kitaura} {et~al.}(2006){Kitaura}, {Janka}, \&
  {Hillebrandt}}]{2006A&A...450..345K}
{Kitaura}, F.~S., {Janka}, H., \& {Hillebrandt}, W. 2006, \aap, 450, 345

\bibitem[{{Lorimer} {et~al.}(2006){Lorimer}, {Stairs}, {Freire}, {Cordes},
  {Camilo}, {Faulkner}, {Lyne}, {Nice}, {Ransom}, {Arzoumanian}, {Manchester},
  {Champion}, {van Leeuwen}, {Mclaughlin}, {Ramachandran}, {Hessels},
  {Vlemmings}, {Deshpande}, {Bhat}, {Chatterjee}, {Han}, {Gaensler}, {Kasian},
  {Deneva}, {Reid}, {Lazio}, {Kaspi}, {Crawford}, {Lommen}, {Backer}, {Kramer},
  {Stappers}, {Hobbs}, {Possenti}, {D'Amico}, \&
  {Burgay}}]{2006ApJ...640..428L}
{Lorimer}, D.~R., {Stairs}, I.~H., {Freire}, P.~C., {et~al.} 2006, \apj, 640,
  428

\bibitem[{{Mandel} \& {O'Shaughnessy}(2010)}]{2010CQGra..27k4007M}
{Mandel}, I. \& {O'Shaughnessy}, R. 2010, Classical and Quantum Gravity, 27,
  114007

\bibitem[{{Nelemans} \& {van den Heuvel}(2001)}]{2001A&A...376..950N}
{Nelemans}, G. \& {van den Heuvel}, E.~P.~J. 2001, \aap, 376, 950

\bibitem[{{Peters}(1964)}]{1964PhRv..136.1224P}
{Peters}, P.~C. 1964, Physical Review, 136, 1224

\bibitem[{{Peters} \& {Mathews}(1963)}]{1963PhRv..131..435P}
{Peters}, P.~C. \& {Mathews}, J. 1963, Physical Review, 131, 435

\bibitem[{{Pfahl} {et~al.}(2005){Pfahl}, {Podsiadlowski}, \&
  {Rappaport}}]{2005ApJ...628..343P}
{Pfahl}, E., {Podsiadlowski}, P., \& {Rappaport}, S. 2005, \apj, 628, 343

\bibitem[{{Schnittman}(2004)}]{2004PhRvD..70l4020S}
{Schnittman}, J.~D. 2004, \prd, 70, 124020

\bibitem[{{Seto} {et~al.}(2001){Seto}, {Kawamura}, \&
  {Nakamura}}]{2001PhRvL..87v1103S}
{Seto}, N., {Kawamura}, S., \& {Nakamura}, T. 2001, Physical Review Letters,
  87, 221103

\bibitem[{{Shapiro Key} \& {Cornish}(2010)}]{2010arXiv1006.3759S}
{Shapiro Key}, J. \& {Cornish}, N.~J. 2010, ArXiv e-prints

\bibitem[{{Sipior} \& {Sigurdsson}(2002)}]{2002ApJ...572..962S}
{Sipior}, M.~S. \& {Sigurdsson}, S. 2002, \apj, 572, 962

\bibitem[{{Smith} \& {LIGO Scientific
  Collaboration}(2009)}]{2009CQGra..26k4013S}
{Smith}, J.~R. \& {LIGO Scientific Collaboration}. 2009, Classical and Quantum
  Gravity, 26, 114013

\bibitem[{{Spallicci} {et~al.}(2005){Spallicci}, {Aoudia}, {de Freitas
  Pacheco}, {Regimbau}, \& {Frossati}}]{2005CQGra..22S.461S}
{Spallicci}, A.~D.~A.~M., {Aoudia}, S., {de Freitas Pacheco}, J., {Regimbau},
  T., \& {Frossati}, G. 2005, Classical and Quantum Gravity, 22, 461

\bibitem[{{Taam} \& {Sandquist}(2000)}]{2000ARA&A..38..113T}
{Taam}, R.~E. \& {Sandquist}, E.~L. 2000, \araa, 38, 113

\bibitem[{{Timmes} {et~al.}(1996){Timmes}, {Woosley}, \&
  {Weaver}}]{1996ApJ...457..834T}
{Timmes}, F.~X., {Woosley}, S.~E., \& {Weaver}, T.~A. 1996, \apj, 457, 834

\bibitem[{{Van Den Broeck}(2010)}]{2010arXiv1003.1386V}
{Van Den Broeck}, C. 2010, ArXiv e-prints

\bibitem[{{Voss} \& {Tauris}(2003)}]{2003MNRAS.342.1169V}
{Voss}, R. \& {Tauris}, T.~M. 2003, \mnras, 342, 1169

\bibitem[{{Webbink}(1984)}]{1984ApJ...277..355W}
{Webbink}, R.~F. 1984, \apj, 277, 355

\bibitem[{{Weisberg} \& {Taylor}(2005)}]{2005ASPC..328...25W}
{Weisberg}, J.~M. \& {Taylor}, J.~H. 2005, in Astronomical Society of the
  Pacific Conference Series, Vol. 328, Binary Radio Pulsars, ed. {F.~A.~Rasio
  \& I.~H.~Stairs}, 25--+

\bibitem[{{Wolszczan}(1991)}]{1991Natur.350..688W}
{Wolszczan}, A. 1991, \nat, 350, 688

\end{thebibliography}

\end{document}